\documentclass[oldversion]{aa}

\usepackage{graphicx,amsmath}
\usepackage{amssymb}
\usepackage{color}
\usepackage{natbib}             
\usepackage{url}
\usepackage{multirow}
\usepackage{threeparttable}
\usepackage{soul,mathabx}
\bibpunct{(}{)}{;}{a}{}{,} 

\begin{document}

   \title{The polar orbit of the warm Neptune GJ\,436b seen with VLT/ESPRESSO\thanks{Based in part on Guaranteed Time Observations collected at the European Southern Observatory under ESO programme 1102.C-0744 by the ESPRESSO Consortium}}
                
   \author{    
V.~Bourrier\inst{\ref{inst:1}},
M.~R.~Zapatero~Osorio\inst{\ref{inst:2}},
R.~Allart\inst{\ref{inst:14},\ref{inst:1}},
O.~Attia\inst{\ref{inst:1}}, 
M.~Cretignier\inst{\ref{inst:1}},  
X.~Dumusque\inst{\ref{inst:1}}, 
C.~Lovis\inst{\ref{inst:1}},
V.~Adibekyan\inst{\ref{inst:3},\ref{inst:10}}, 
F.~Borsa\inst{\ref{inst:8}},
P.~Figueira\inst{\ref{inst:11},\ref{inst:3}},  
J.I.~Gonz\'alez~Hern\'andez\inst{\ref{inst:5},\ref{inst:6}},   
A.~Mehner\inst{\ref{inst:15}},
N.~C.~Santos\inst{\ref{inst:3},\ref{inst:10}},
T.~Schmidt\inst{\ref{inst:1}},
J.~V.~Seidel\inst{\ref{inst:1},\ref{inst:17}},
A.~Sozzetti\inst{\ref{inst:13}},
Y.~Alibert\inst{\ref{inst:16}}, 
N.~Casasayas-Barris\inst{\ref{inst:9}},
D.~Ehrenreich\inst{\ref{inst:1}},       
G.~Lo~Curto\inst{\ref{inst:15}},  
C.~J.~A.~P.~Martins\inst{\ref{inst:3},\ref{inst:4}},     
P.~Di~Marcantonio\inst{\ref{inst:7}},  
D.~M\'egevand\inst{\ref{inst:1}},       
N.J.~Nunes\inst{\ref{inst:12}},
E.~Palle\inst{\ref{inst:5},\ref{inst:6}},
E.~Poretti\inst{\ref{inst:8}},  
S.~G.~Sousa\inst{\ref{inst:3}}}

\authorrunning{V.~Bourrier et al.}
\titlerunning{Refined architecture of the GJ\,436 system}

\offprints{V.B. (\email{vincent.bourrier@unige.ch})}

   \institute{Observatoire Astronomique de l'Universit\'e de Gen\`eve, Chemin Pegasi 51b, CH-1290 Versoix, Switzerland\label{inst:1} \and
Centro de Astrobiología (CSIC-INTA), Carretera de Ajalvir km 4, 28850 Torrejón de Ardoz, Madrid, Spain  \label{inst:2}    \and
Instituto de Astrof\'{\i}sica e Ci\^encias do Espa\c co, CAUP, Universidade do Porto, Rua das Estrelas, 4150-762, Porto, Portugal \label{inst:3}    \and
Centro de Astrof\'{\i}sica da Universidade do Porto, Rua das Estrelas, 4150-762 Porto, Portugal \label{inst:4}    \and
Instituto de Astrof\'{\i}sica de Canarias, Via Lactea, E-38200 La Laguna, Tenerife, Spain \label{inst:5}    \and
Universidad de La Laguna, Departamento de Astrof\'isica, E- 38206, La Laguna, Tenerife, Spain \label{inst:6}    \and
INAF, Osservatorio Astronomico di Trieste, via G. B. Tiepolo 11, I-34143, Trieste, Italy \label{inst:7}    \and
INAF, Osservatorio Astronomico di Brera, Via Bianchi 46, 23807 Merate, Italy \label{inst:8}    \and
Leiden Observatory, Leiden University, Postbus 9513, 2300 RA Leiden, The Netherlands \label{inst:9}    \and
Departamento de F\'{\i}sica e Astronomia, Faculdade de Ci\^encias, Universidade do Porto, Rua do Campo Alegre, 4169-007 Porto, Portugal \label{inst:10}    \and
European Southern Observatory, Alonso de Cordova, Vitacura, Santiago, Chile \label{inst:11}    \and
Instituto de Astrof\'{\i}sica e Ci\^encias do Espa\c co, Faculdade de Ci\^encias da Universidade de Lisboa, Edif\'icio C8, Campo Grande, PT1749-016
Lisbon, Portugal \label{inst:12}    \and
INAF Osservatorio Astrofisico di Torino, Via Osservatorio 20, 10025 Pino Torinese, Italy \label{inst:13}    \and
Department of Physics, and Institute for Research on Exoplanets, Universit\'e de Montr\'eal, Montr\'eal, H3T 1J4, Canada \label{inst:14}    \and
European Southern Observatory, Karl-Schwarzschild-Str. 2, 85748 Garching bei M\"unchen, Germany \label{inst:15}    \and
Physics Institute of University of Bern, Gesellschaftsstrasse 6, CH-3012 Bern, Switzerland \label{inst:16}    \and
European Southern Observatory, Alonso de C\'ordova 3107, Vitacura, Casilla 190001, Santiago, Chile  \label{inst:17}   
}

   
   \date{} 
 
  \abstract
{GJ\,436b might be the prototype of warm Neptunes that have undergone late migration induced by an outer companion. Precise determination of the orbital architecture of such systems is critical to constraining their dynamical history and evaluating the role of delayed migration in the exoplanet population. To this purpose we analyzed the Rossiter-McLaughlin (RM) signal of GJ\,436 b in two transits ---recently observed with ESPRESSO--- using three different techniques. The high level of precision achieved in radial velocity (RV) measurements allows us to detect the deviation from the Keplerian orbit, despite the slow rotation of the M dwarf host ($v$\,sin\,$i_*$ = 272.0$\stackrel{+40.0}{_{-34.0}}$\,m\,s$^{-1}$), and to measure the sky-projected obliquity ($\lambda$ = 102.5$\stackrel{+17.2}{_{-18.5}}^{\circ}$). The Reloaded RM technique, which allows the stellar RV field along the transit chord to be analyzed, yields $\lambda$ = 107.5$\stackrel{+23.6}{_{-19.3}}^{\circ}$ and $v$\,sin\,$i_*$ = 292.9$\stackrel{+41.9}{_{-49.9}}$\,m\,s$^{-1}$. The RM Revolutions technique, which allows us to fit the spectral profiles from all planet-occulted regions together, yields $\lambda$ = 114.1$\stackrel{+22.8}{_{-17.8}}^{\circ}$ and $v$\,sin\,$i_*$ = 300.5$\stackrel{+45.9}{_{-57.0}}$\,m\,s$^{-1}$. The consistent results between these three techniques, and with published results from HARPS/HARPS-N data, confirm the polar orbit of GJ\,436b and support the hypothesis that its origin lies in Kozai migration. Results from a joint RM Revolutions analysis of the ESPRESSO, HARPS, and HARPS-N datasets ($\lambda$ = 113.5$\stackrel{+23.3}{_{-17.3}}^{\circ}$; $v$\,sin\,$i_*$ = 293.5$\stackrel{+43.7}{_{-52.2}}$\,m\,s$^{-1}$) combined with a revised stellar inclination ($i_\star$ = 35.7$\stackrel{+5.9}{_{-7.6}}^{\circ}$ or 144.2$\stackrel{+7.6}{_{-5.9}}^{\circ}$) lead us to constrain the 3D obliquity $\Psi$ to 103.2$\stackrel{+12.8}{_{-11.5}}^{\circ}$. }


\maketitle

\section{Introduction}
\label{intro} 

GJ\,436 was the first M dwarf across which the Rossiter-McLaughlin (RM) effect was detected, leading to the measurement of a highly misaligned orbit (\citealt{Bourrier_2018_Nat}) for its evaporating (\citealt{Kulow2014,Ehrenreich2015,Lavie2017,DosSantos2019}) warm Neptune companion. This peculiar architecture supports a late Kozai migration (\citealt{Naoz2016}) induced by an outer companion (\citealt{Bourrier_2018_Nat}), a scenario that was originally proposed by \citet{Beust2012} to explain the current nonzero orbital eccentricity of GJ\,436b ($e$ = 0.152, \citealt{Trifonov2018}), which should otherwise have been quickly circularized by tidal interactions with the star (\citealt{Mardling2008}). In a Kozai scenario, a distant perturbing body traps the inner planet in a secular dynamical resonance far away from the star for several billion years, only to subsequently decouple and migrate inwards on a misaligned and eccentric orbit (e.g., \citealt{Fabrycky2007}, \citealt{Anderson2016}). GJ\,436b might be a representative example of Neptune-size planets that started to evaporate long after their formation because of this delayed migration (\citealt{Bourrier_2018_Nat,Owen2018,Correia2020,Attia2021}). \citealt{Albrecht2021} recently presented statistical evidence for a preponderance of exoplanets on nearly polar orbits, several of which have noncircular orbits, lie at the edge of the hot Neptune desert, and are observed to be evaporating. Noticeable examples with known planetary companions include HAT-P-11\,b (see \citealt{Yee2018,BenJaffel2021}) and WASP-107\,b (see \citealt{Spake2018,Allart2019,Kirk2020,Piaulet2021,Spake2021}). GJ\,3470\,b is also worth mentioning as it exhibits all of the aforementioned characteristics of a delayed Kozai migration (see \citealt{Bourrier2018_GJ3470b,Palle2020,Ninan2020,Stefansson2021}), although, like GJ\,436b, its putative companion remains undetected. While secular resonance crossing driven by the protoplanetary disk (\citealt{Petrovich2020}) can also account for the eccentric and misaligned orbits of low-mass planets around the desert, it requires that the planet formed on a close-in orbit and thus does not explain the long-term survival of their evaporating atmospheres. The above systems thus support the critical role that Kozai resonance could play in forming the desert of hot Neptunes, by delaying the arrival at their close-in location and their subsequent evaporation. The case of GJ\,436b illustrates the utility of transit spectroscopy in determining the spin--orbit angles of close-in planets (see the review by \citealt{Triaud2018}), in constraining the mechanisms that altered their orbital architecture, and in unraveling their complex evolution.

Here we analyze the RM effect of GJ\,436b in two transits recently observed with ESPRESSO (Sect.~\ref{sec:data}). We compare the results obtained using the classical analysis of the RV anomaly (Sect.~\ref{sec:DI_CCFs}) with those obtained using the recently developed Reloaded RM and RM Revolutions techniques, analyzing the stellar surface properties along the transit chord (Sect.~\ref{sec:Loc_CCFs}). Using the latter technique, we perform a joint fit to the ESPRESSO and archival HARPS/HARPS-N data to refine the architecture of the GJ\,436 system (Sect.~\ref{sec:glob_fit}).


\section{Observations and data reduction}
\label{sec:data}

We exploit two datasets obtained with ESPRESSO during the transit of GJ\,436b on 27 February 2019 (Visit 1) and 29 April 2019  (Visit 2). During each visit, 49 exposures were obtained at about the same orbital phases, with 10 in-transit exposures, 17 before the transit in Visit 1 (respectively 16 in Visit 2), and 22 after the transit in Visit 1 (respectively 23 in Visit 2). GJ\,436 was observed on fiber A, while fiber B was used to monitor sky contamination. Because of the faintness of the star in the optical (V = 10.6), it was observed in ESPRESSO high-resolution mode (R$\sim$140000) with a binning of 2x1 to reduce readout noise (instrumental mode HR21, for more information see \citealt{Pepe2021}). Spectra were extracted from the detector images, corrected, and calibrated using version 2.2.8 of the Data Reduction Software (DRS) pipeline (\citealt{Pepe2021}). One of the DRS corrections concerns the color effect caused by the variability of extinction induced by Earth's atmosphere (e.g., \citealt{bourrier2014b}, \citealt{Bourrier_2018_Nat}, \citealt{Wehbe2020}). The flux balance of the GJ\,436 spectra was reset to a M3 stellar spectrum template before they were passed through weighted cross-correlation (\citealt{baranne1996}; \citealt{pepe2002}) with an M3 numerical mask to compute cross-correlation functions (CCFs) with a step of 0.5\,km\,s$^{-1}$. This mask is part of a new set that were built with weights more representative of the photonic error on the line positions, as described in \citet{Bourrier2021}.

Properties of the GJ\,436 system used and derived in our analyses are reported in Table~\ref{tab:sys_prop}. Recent CHEOPS observations of GJ\,436b (\citealt{Maxted2021}) improve the precision on mid-transit times at the epochs of our observations to 8\,s, removing potential biases associated with ephemeris (e.g., \citealt{CasasayasBarris2021}). The planet-to-star radius ratio was fixed to the error-weighted average of values obtained at different precisions with various instruments by \citet{Lanotte2014}, \citet{Knutson2014}, \citet{Lothringer2018}, and \citet{Maxted2021}. None of these studies found any strong evidence for variations in transit depth with wavelength.

Throughout the paper, the posterior probability distributions (PDFs) of free parameters describing models fitted to the data are sampled using \textit{emcee} MCMC (\citealt{Foreman2013}), adjusting the number of walkers and the burn-in phase based on the degrees of freedom of the considered problem and the convergence of the chains. Best-fit values for these parameters are set to the median of their PDFs, and their 1$\sigma$ uncertainty ranges are defined using highest density intervals.

\begin{table}[tbh]
\caption{Properties of the GJ\,436 system.}\centering
\begin{threeparttable}
\begin{tabular}{lcl}
\hline
\noalign{\smallskip}  
\textbf{Parameter}                &       \textbf{Value}  & \textbf{Unit}          \\     
\noalign{\smallskip}
\hline
\hline
\noalign{\smallskip} 
Stellar radius    &  0.425$^\mathrm{[1]}$     & $R_{\Sun}$   \\
Projected rotational velocity  &   293.5$\stackrel{+43.7}{_{-52.2}}^{\dagger}$  & m\,s$^{-1}$ \\                                                                                              
Stellar inclination  &   35.7$\stackrel{+5.9}{_{-7.6}}^{\dagger}$  & deg  \\  
                                         &   144.2$\stackrel{+7.6}{_{-5.9}}^{\dagger}$  & deg  \\  
Power-2 limb-darkening & $c_\mathrm{1}$ = 0.900$^\mathrm{[1]}$    &     \\
                       & $c_\mathrm{2}$ =  0.508$^\mathrm{[1]}$   &     \\                                                                                                    
\hline
Stellar reflex motion amplitude  &  17.38$^\mathrm{[2]}$  &   m\,s$^{-1}$                                                                                               \\
Orbital period &   2.64389803$^\mathrm{[3]}$   & days  \\
Eccentricity    &   0.152$^\mathrm{[2]}$  &  \\
Argument of periastron    &   325.8$^\mathrm{[2]}$   & deg  \\  
Mid-transit time (-2~450~000)   & 5475.82450$^\mathrm{[1]}$   & BJD$_{\rm TDB}$  \\
Orbital inclination     & 86.7889$^\mathrm{[1]}$   & deg   \\
Scaled semi-major axis     &  14.46$^\mathrm{[1]}$  & \\  
Planet-to-star radius ratio   &  0.08315$^\mathrm{[1]}$    &  \\
Projected spin--orbit angle  &  113.5$\stackrel{+23.3}{_{-17.3}}^{\dagger}$  & deg  \\
3D spin--orbit angle   &  103.2$\stackrel{+12.8}{_{-11.5}}^{\dagger}$  & deg  \\
\hline
\end{tabular}
\begin{tablenotes}[para,flushleft]
Notes. Parameters from [1] \citet{Maxted2021}, [2] \citet{Trifonov2018}, [3] \citet{Lanotte2014} are fixed. Parameters with ${\dagger}$ are derived from our final analysis in Sect.~\ref{sec:glob_fit}. For details about the power-2 limb-darkening law, see \citet{Morello2017,Maxted2018}. The 3D spin--orbit angle is combined over the two degenerate configurations for the stellar inclination, see text.
  \end{tablenotes}
  \end{threeparttable}
\label{tab:sys_prop}
\end{table}


\section{Analysis of the disk-integrated CCFs}
\label{sec:DI_CCFs}

\subsection{Stellar line properties}
\label{sec:DI_fit}

\begin{figure}
\includegraphics[trim=1cm 0cm 0cm 0cm,clip=true,width=\columnwidth]{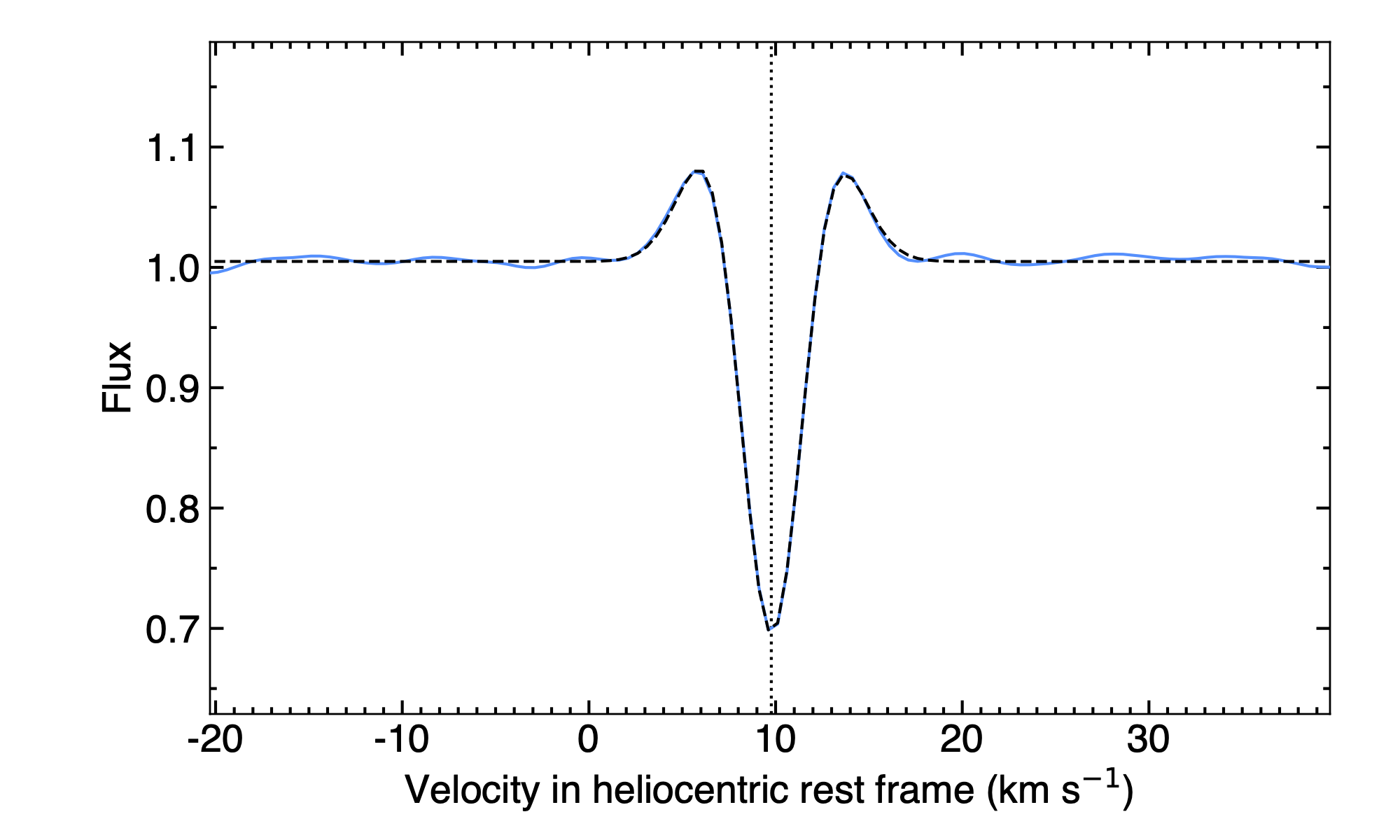}
\centering
\caption[]{Master-out CCF$_\mathrm{DI}$ of GJ\,436 in Visit 1 (blue profile) and its best-fit double-Gaussian model (black dashed profile). The vertical dotted line indicates the measured systemic velocity.}
\label{fig:mout_V1}
\end{figure}

The CCF$_\mathrm{DI}$ (for ``disk-integrated'') produced by the DRS correspond to the light coming from the entire star. CCF$_\mathrm{DI}$ of each individual exposure are aligned by correcting their velocity table for the Keplerian motion of the star. CCF$_\mathrm{DI}$ outside of the transits are co-added to build master-outs representative of the unocculted star, which display side lobes typical of M dwarf CCFs (Fig.~\ref{fig:mout_V1}). The lobes are thought to arise from the region around each mask line devoid of other stellar lines, while the lower continuum beyond the lobes would arise from the dense forest of lines in these cool star spectra being well-characterized by an average distance between lines. As such, the peaks of the lobes are better estimates of the actual continuum. We use the double-Gaussian model introduced by \citet{Bourrier_2018_Nat} to fit the CCFs of GJ\,436, as it captures their full profile  well and was shown to improve the stability of the measurements. The model consists of the sum of a Gaussian function representing the continuum and side lobes, and an inverted Gaussian function representing the core. Model parameters are the amplitude, FWHM, and RV centroid of the core Gaussian component, and the amplitude ratio, FWHM ratio, and RV centroid difference between the lobe and core Gaussian components.

All CCF$_\mathrm{DI}$ are shifted to the rest frame of the  star using the centroids derived from the fits to the master-outs. The best-fit double-Gaussian model to the master-out CCF$_\mathrm{DI}$ in each visit is then used as a template (i.e., the amplitude ratio, FWHM ratio, and RV centroid difference between the lobe and core Gaussian components are fixed to its properties) to fit each individual exposure. We performed the same operations on CCFs reduced with the mask for M3-type stars previously used by the ESPRESSO DRS, and used the out-of-transit time-series of RV residuals, contrast, and FWHM derived from the fits to both CCF series to assess the stability of the measurements and the improvements associated with the new CCF mask (Table~\ref{tab:disp_full}). This mask yields shallower but much narrower CCF$_\mathrm{DI}$, with relative precisions on the contrast and FWHM improved by 27\%\ and 44\% in Visit 1, 8\%\ and 39\% in Visit 2, respectively. The stability of the RV measurements from Visits 1 and 2 is improved by 8\%\ and 19\%, respectively. This analysis suggests that the new masks devised in \citet{Bourrier2021} improve the stability of CCF not only from K-type stellar spectra but also from M dwarf spectra. The above comparison and subsequent analyses were performed with CCF$_\mathrm{DI}$ reduced by the DRS from the sky-subtracted products, as this improves the overall stability of their properties despite an increased dispersion on the contrast in Visit 2.

\begin{table}[h]
\centering
\caption{Quality assessment of ESPRESSO CCF$_\mathrm{DI}$}
\begin{tabular}{lcccc}
\hline
Sky correction   &     Yes & No & Yes \\
\hline
Mask &            Classic & Improved & Improved \\
\hline
\hline
$<$Contrast$>$ (\%)      & 33.993 &   30.677   & 30.723  \\
                                                  &  33.958  &  30.625    &  30.674 \\
e$_\mathrm{Contrast}$ (\%)      &  0.012  &  0.013    &  0.014 \\
                                                           &  0.011  &  0.012    &  0.013 \\
$\sigma_\mathrm{Contrast}$ (\%)    &  0.024  &    0.016  &  0.016 \\
                                                  &  0.021  &   0.015   &  0.017 \\         
$\sigma_\mathrm{Contrast}^\mathrm{rel}$ (ppm)    & 697   &  523    &  507 \\
                                                                                        &  605  &    495  & 558  \\                                         
\hline
$<$FWHM$>$ (km\,s$^{-1}$)   & 7.0304   &   4.9606   &  4.9577 \\
                                                  &  7.0360  &   4.9641   &  4.9652 \\
e$_\mathrm{FWHM}$ (m\,s$^{-1}$)   & 2.2   &  1.8    &  1.9 \\
                                                  &  2.0  &  1.7    &  1.7 \\                
$\sigma_\mathrm{FWHM}$ (m\,s$^{-1}$)   &  3.7  &  1.6    & 1.5  \\
                                                  &  2.9  &   1.3   & 1.2  \\ 
$\sigma_\mathrm{FWHM}^\mathrm{rel}$ (ppm)    &  527  &   328   &  296 \\
                                                                                         & 412   &    259  &  252 \\                                       
\hline
e$_\mathrm{RV}$ (cm\,s$^{-1}$)   & 78   &  67    &  71 \\
                                                                &  69  &  61    &  63 \\  
$\sigma_\mathrm{RV}$ (cm\,s$^{-1}$)    &  78  &   70   &  72 \\
                                                  &  72  &  64    &  58 \\                                        
\hline
\end{tabular}
\begin{tablenotes}[para,flushleft]
Notes: For each property, values on the first (resp.) line correspond to Visit 1 (resp. 2). $\sigma$ indicates standard deviations of the out-of-transit contrast, FWHM, and Keplerian RV residuals with respect to their mean value ($<$x$>$). The $\sigma^\mathrm{rel}$ dispersions have been normalized by this mean to allow for a direct comparison between cases. e$_\mathrm{x}$ indicate the mean out-of-transit error on $x$.
\end{tablenotes}
\label{tab:disp_full}
\end{table}


\subsection{Classical RM analysis}
\label{sec:class_RM}

\begin{figure}
\includegraphics[trim=1cm 1cm 1.5cm 2.1cm,clip=true,width=\columnwidth]{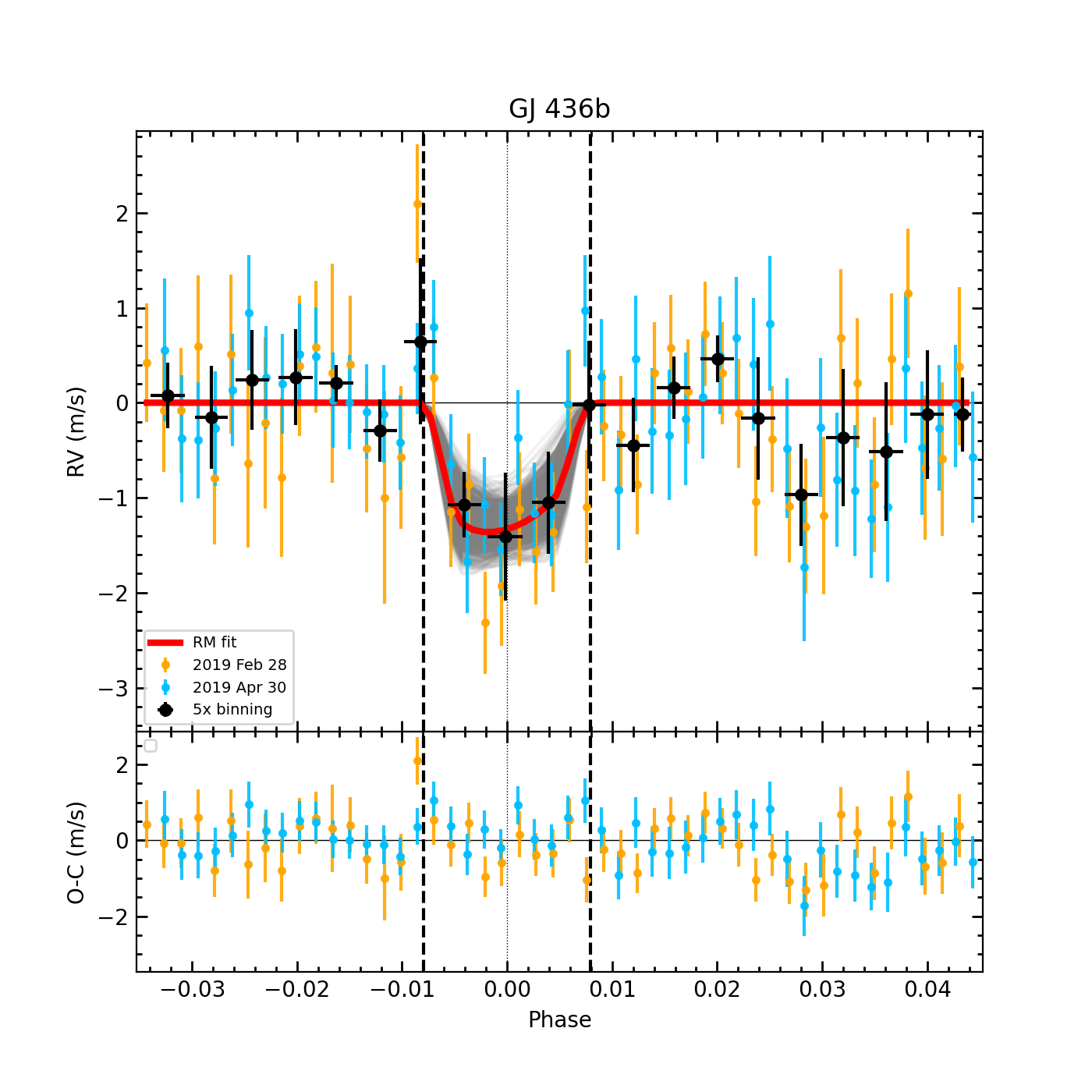}
\centering
\caption[]{\textit{Top}: RM anomaly induced by GJ\,436\,b in the RV residuals measured in Visits 1 (orange) and 2 (blue). Black dots represent a binning of the RV data with a bin size of 2.5 over the combined transits. The red line shows the classical fit to the anomaly, with associated uncertainty depicted by the gray area. Vertical dashed lines indicate the first and fourth transit contacts. \textit{Bottom}: RV residuals, which display an overall dispersion of 65\,cm\,s$^{-1}$.}
\label{fig:arome}
\end{figure}

Figure~\ref{fig:arome} shows the RV time-series derived from the CCF$_\mathrm{DI}$. In contrast to the HARPS/HARPS-N data in \citet{Bourrier_2018_Nat} the precision of the ESPRESSO data is sufficient to detect the anomaly induced by the planet, despite its small amplitude ($\sim$1\,m\,s$^{-1}$). The RM signal is consistent between the two visits within the quoted error bars, strengthening its velocimetric detection. We fitted the two visits together using the ARoME code\footnote{\url{http://www.astro.up.pt/resources/arome}}, which implements the prescription of \citet{boue2013}. Residuals on RV were phase-folded using the revised ephemeris (Table~\ref{tab:sys_prop}). The ARoME code does not model the Power-2 limb-darkening law used by \citet{Maxted2021}, and we therefore used the limb-darkening calculator tool\footnote{\url{https://exoctk.stsci.edu/limb\_darkening}. This tool is part of the Exoplanet Characterization toolkit, \url{https://exoctk.readthedocs.io/en/latest/}} to derive quadratic coefficients $c1 = 0.440 \pm 0.008$ and $c2 = 0.238 \pm 0.012$ in the wavelength range of ESPRESSO (adopting the Phoenix ACES model atmospheres; \citealt{Husser2013}). We also adopted a stellar macroturbulence velocity of 1 km\,s$^{-1}$ from Fig.~11 in \citet{Wende2009} (although this parameter has little impact in our final solution), the planetary parameters reported in Table~\ref{tab:sys_prop}, and an ESPRESSO instrumental CCF width of 2.1\,km\,s$^{-1}$ \citep{Pepe2021}. 

We ran a MCMC simulation with 4000 samples in order to derive the sky-projected obliquity $\lambda$ and stellar rotational velocity $v$\,sin\,$i_{\star}$, together with their associated uncertainties. We set wide and uniform priors, $\mathcal{U}$(0,180)\,$^{\circ}$ on $\lambda$ and $\mathcal{U}$(0, 5)\,km\,s$^{-1}$ on $v$\,sin\,$i$. The best-fit RM model shown in Figure~\ref{fig:arome} is obtained for $\lambda = 102.5^{+17.2}_{-18.5}$\,$^{\circ}$ and $v$\,sin\,$i_{\star} = 272^{+40}_{-34}$\,m\,s$^{-1}$. These results are consistent within 1$\sigma$ with the values reported in \cite{Bourrier_2018_Nat}, although their uncertainties are reduced by up to a factor of two given the higher S/N and larger spectral coverage of the ESPRESSO data (Table~\ref{tab:RM_results}).

\begin{table*}[h]
\begin{minipage}[h!]{\textwidth}
\centering
\caption{Comparison of RM analyses.}
\begin{tabular}{lccc}
\hline
        Instruments     &  Analysis      &   \multicolumn{2}{c}{Parameters}  \\
\hline
        &                &                                               $v$\,sin\,$i_{\star}$ (m\,s$^{-1}$)  & $\lambda$ ($^{\circ}$)            \\
\hline
\hline
ESPRESSO  &  Classical RM                       &   272.0$\stackrel{+40.0}{_{-34.0}}$ &   102.5$\stackrel{+17.2}{_{-18.5}}$    \\
ESPRESSO    & Reloaded RM                       & 292.9$\stackrel{+41.9}{_{-49.9}}$ &   107.5$\stackrel{+23.6}{_{-19.3}}$    \\
ESPRESSO    &  RM Revolutions                   &  300.5$\stackrel{+45.9}{_{-57.0}}$ &   114.1$\stackrel{+22.8}{_{-17.8}}$    \\
 HARPS/HARPS-N      & \cite{Bourrier_2018_Nat}   &  330$\stackrel{+91}{_{-66}}$ &   72$\stackrel{+33}{_{-24}}$    \\
 HARPS/HARPS-N      &  RM Revolutions  &  290.1$\stackrel{+127.7}{_{-147.9}}$ &   102.4$\stackrel{+57.4}{_{-50.4}}$    \\
HARPS/HARPS-N/ESPRESSO    & RM Revolutions                      & 293.5$\stackrel{+43.7}{_{-52.2}}$ &   113.5$\stackrel{+23.3}{_{-17.3}}$    \\                               
\hline
\end{tabular}
\begin{tablenotes}[para,flushleft]
Notes: The analysis from \cite{Bourrier_2018_Nat} was performed with an older version of the Reloaded RM technique and a different reduction of the CCFs.
\end{tablenotes}
\label{tab:RM_results}
\end{minipage}
\end{table*}


\section{Analysis of the planet-occulted CCFs}
\label{sec:Loc_CCFs}

RVs condense the information contained in disk-integrated CCF profiles into a single measurement, which limits our ability to detect and interpret the occultation of the stellar surface by the planet in the resulting RV anomaly. A classical analysis of the velocimetric RM effect can be  further biased if the occulted stellar line profile is not well modeled or varies along the transit chord (\citealt{Cegla2016a}, \citealt{Bourrier2017_WASP8}). Here we analyze the data using the Reloaded (\citealt{Cegla2016}) and Revolutions (\citealt{Bourrier2021}) techniques, which provide direct access to the occultation signal and avoid these biases through the analysis of the planet-occulted starlight rather than the disk-integrated starlight. In both techniques, the disk-integrated CCFs are rescaled to their correct relative flux level, which is critical to extract CCFs from the planet-occulted regions. Whereas the Reloaded RM technique analyzes the stellar line from each planet-occulted region independently to retrieve and interpret the stellar surface RVs along the transit chord, the RM Revolutions technique exploits all available information by analyzing the planet-occulted CCFs together with a joint model. Hereafter, we characterize planet-occulted stellar lines according to their core RV centroid, and also according to the ``true'' contrast and FWHM calculated using the peaks of their model lobes as an estimate for the continuum.

\subsection{Extraction and analysis of individual exposures}
\label{sec:extra}

\begin{figure*}   
\begin{minipage}[h!]{\textwidth}
\includegraphics[trim=0cm 0cm 0cm 0cm,clip=true,width=\columnwidth]{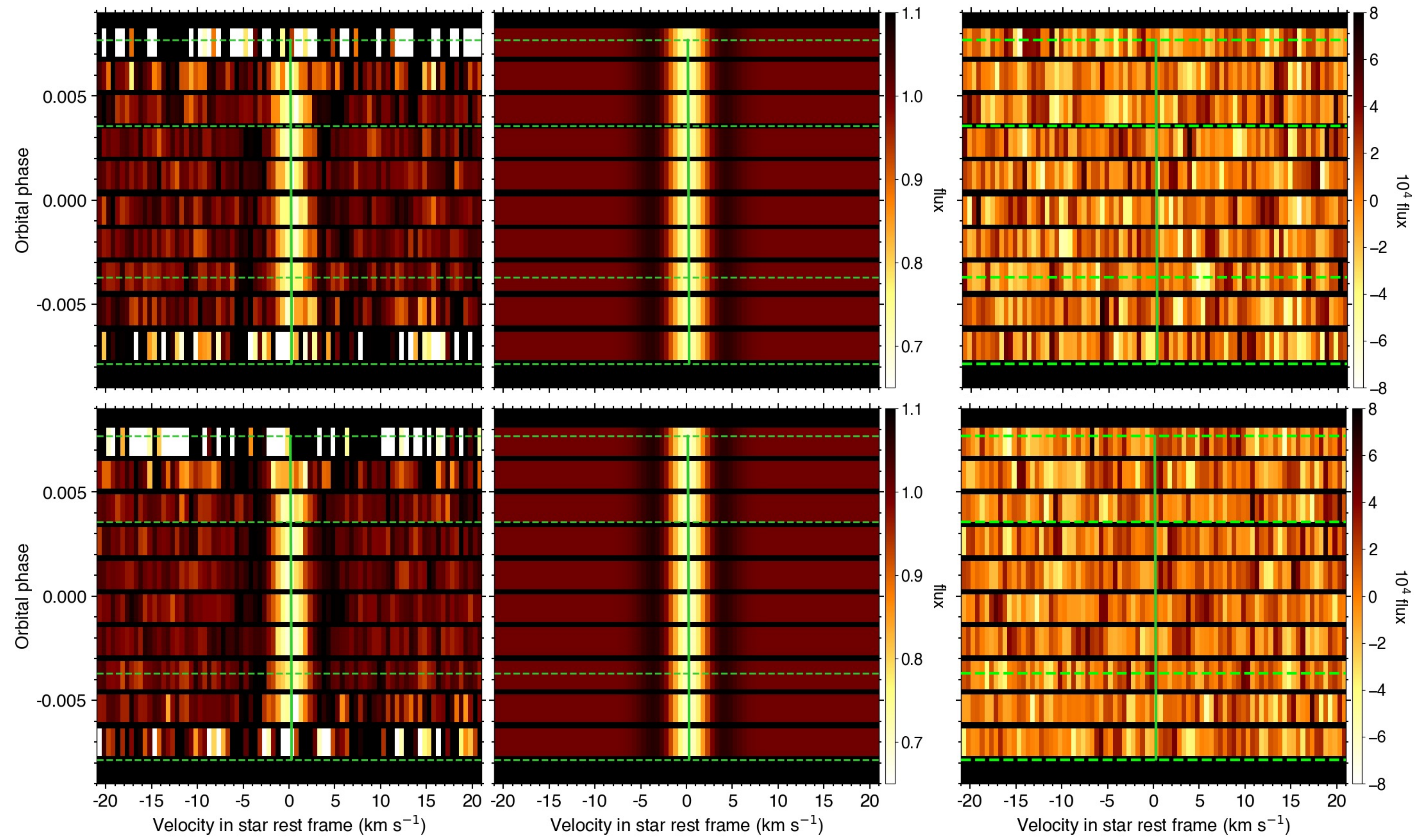}
\centering
\end{minipage}
\caption[]{Maps of the CCF$_\mathrm{intr}$ during the transit of GJ\,436b in Visits 1 (upper panel) and 2 (lower panel). Left panels show measured profiles. The core of the stellar line from the planet-occulted regions is clearly visible as a bright streak, with side lobes visible on both sides as dark streaks. Middle panels show the RM Revolutions best-fit profiles. Right panels show residuals between measured and theoretical profiles. Transit contacts are shown as green dashed lines. Values are colored as a function of the flux, and plotted as a function of RV in the star rest frame (in abscissa) and orbital phase (in ordinate). The green solid line shows the stellar surface RV model from the RM Revolutions best fit.}
\label{fig:CCFintr_map}
\end{figure*}

CCF$_\mathrm{DI}$ are scaled to a common flux level outside of the transit, and to the flux expected from the planetary disk absorption during transit using a light curve computed with the \textit{batman} package (\citealt{Kreidberg2015}) and the properties in Table~\ref{tab:sys_prop}. CCFs from the planet-occulted regions are retrieved by subtracting the scaled CCF$_\mathrm{DI}$ from their corresponding master-out, and are then reset to a common flux level to yield intrinsic CCF$_\mathrm{intr}$ that allow for a more direct comparison of the local stellar lines (Fig.~\ref{fig:CCFintr_map}). 

Following the RM Revolutions approach (\citealt{Bourrier2021}), we first analyze individual exposures to evaluate the quality of each CCF$_\mathrm{intr}$ and assess the possibility for line profile variations along the transit chord. Like the disk-integrated line, the average intrinsic stellar line displays side lobes and CCF$_\mathrm{intr}$ were therefore fitted with a double-Gaussian profile as well. As in \citet{Bourrier_2018_Nat}, we fixed the core-to-lobe FWHM ratio, amplitude ratio, and centroid difference of the CCF$_\mathrm{intr}$ model to the values derived from the fit to the master-out CCF$_\mathrm{DI}$ in each visit. We set uniform priors, $\mathcal{U}$(-1,1)\,km\,s$^{-1}$ on the model RV centroid (\textit{i.e.}, about three times the maximum stellar surface RV), $\mathcal{U}$(0,10)\,km\,s$^{-1}$ on the FWHM (\textit{i.e.}, about three times the width of the CCF$_\mathrm{DI}$, which is assumed to be similar to that of the CCF$_\mathrm{intr}$ given the slow stellar rotation), and $\mathcal{U}$(0,1) on the contrast. We ran a set of 100 walkers for 2000 steps, with a burn-in phase of 500 steps. The model is fitted to the data over [-50,50]\,km\,s$^{-1}$ in the star rest frame. 

\begin{figure}
\includegraphics[trim=0cm 0cm 0cm 0cm,clip=true,width=\columnwidth]{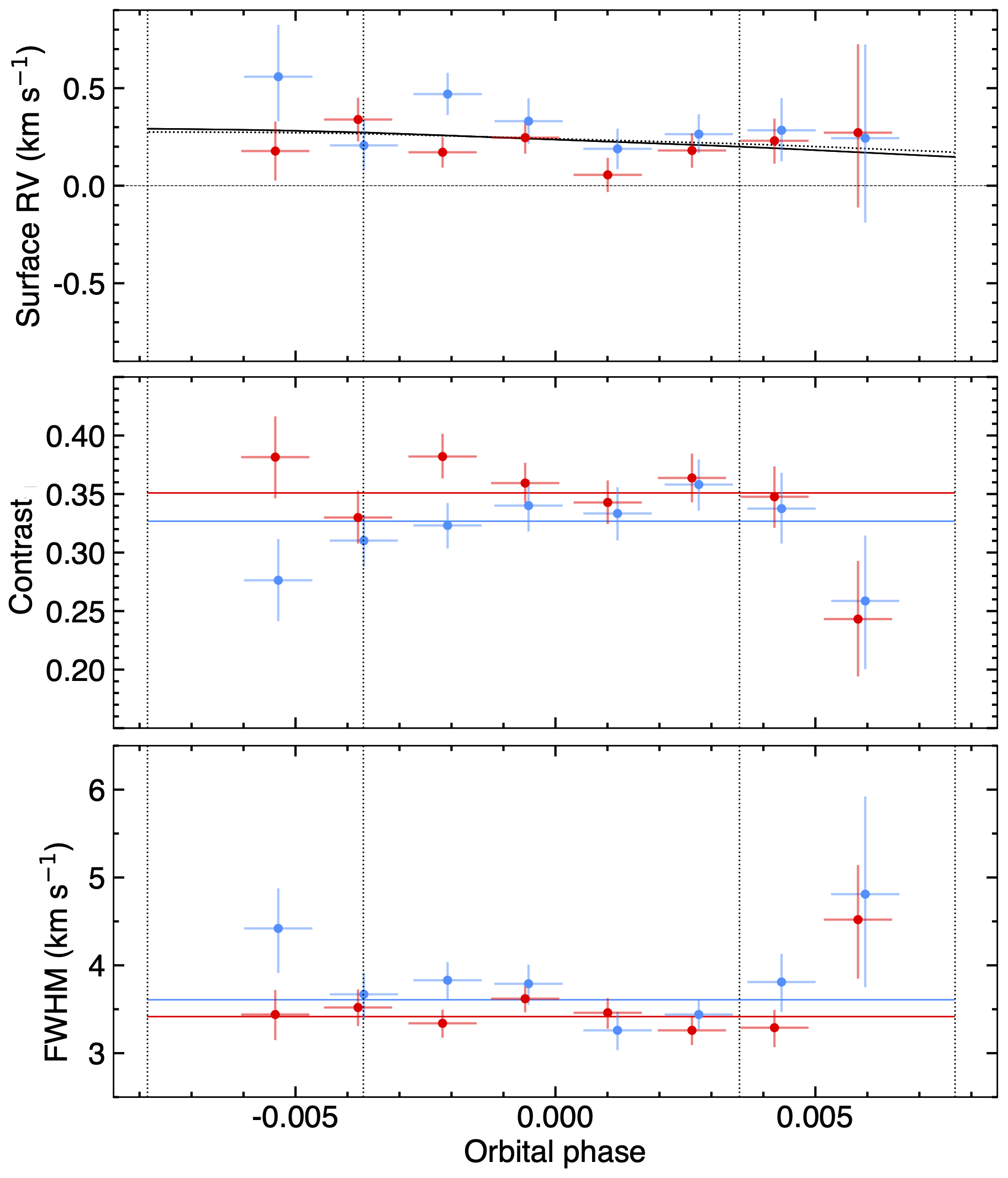}
\centering
\caption[]{Properties of the stellar surface regions occulted by GJ\,436b in ESPRESSO Visits 1 (blue) and 2 (red), derived from the fits to individual exposures. Vertical bars indicate their associated 1$\sigma$ highest density intervals. Horizontal bars indicate the duration of each exposure. Solid lines are derived from the best RM Revolutions fit to the CCF$_\mathrm{intr}$ maps (contrast and FWHM are specific to each visit and colored accordingly). The dotted line in the top panel is the best Reloaded RM fit to the surface RV series, nearly identical to the model from the RM Revolutions fit. Dashed vertical lines are the transit contacts.}
\label{fig:PropLoc_GJ436b}
\end{figure}

The stellar line is clearly detected in all individual CCF$_\mathrm{intr}$ (Fig.~\ref{fig:CCFintr_map}), with well-defined PDFs for their model parameters (Fig.~\ref{fig:PDF_ESPRESSO_indivexp}), except for the first and last exposures in both visits. Their S/N is much lower than the rest of the series, which is due to the darkened flux of the stellar limb and its partial occultation by the planet, and they were excluded from further analysis. Properties of the local stellar lines along the transit chord are shown in Fig.~\ref{fig:PropLoc_GJ436b}. The local line contrast, FWHM, and RVs are consistent between the two visits, as expected from the stability of the host star (\citealt{Bourrier_2018_Nat}). Surface RVs are all positive and remain at about the same value between 200 and 300\,m\,s$^{-1}$, which is in agreement with the results from \citet{Bourrier_2018_Nat} and the classical RM analysis (Sect.~\ref{sec:class_RM}), and confirms that GJ\,436b is on a polar orbit across the redshifted stellar hemisphere.


\subsection{Reloaded RM fit to surface RVs}
\label{sec:RMR}

The Reloaded method is restricted to the interpretation of the RV centroids from the planet-occulted CCFs, which requires that they have a strong enough S/N to be fitted with a stellar line model. This is the case here, and we were therefore able to fit the model described in \citet{Cegla2016} to the surface RV series. We assume solid-body rotation for the star and oversample each exposure by 5 to account for the possible blur induced by the planet motion (\citet{Bourrier2017_WASP8}). Uniform priors are set on the model parameters, $\mathcal{U}$(0,10)\,km\,s$^{-1}$ for $v$\,sin\,$i_*$, and over its definition range $\mathcal{U}$(-180,180)$^{\circ}$ for $\lambda$. We ran 20 walkers for 2000 steps, with a burn-in phase of 500 steps. The best fit yields $v$\,sin\,$i_*$ = 292.9$\stackrel{+41.9}{_{-49.9}}$\,m\,s$^{-1}$ and $\lambda$ = 107.5$\stackrel{+23.6}{_{-19.3}}^{\circ}$. These results are consistent within 1$\sigma$ with those derived from the Reloaded analysis of HARPS/HARPS-N transits in \citet{Bourrier_2018_Nat} ($v$\,sin\,$i_*$ = 330$\stackrel{+91}{_{-66}}$\,m\,s$^{-1}$; $\lambda$ = 72$\stackrel{+33}{_{-24}}^{\circ}$), and comparable in terms of precision with the results from the classical RM analysis (Table~\ref{tab:RM_results}). We note that, despite their precision, the ESPRESSO data on GJ\,436b do not allow us to constrain differential rotation, as a fit following the prescription of \citet{Cegla2016} results in a uniform PDF for the relative differential rotation rate. This is in agreement with the conclusions of \citet{Browning2008} and \citet{Morin2008} that differential rotation in fully convective dwarf stars is quenched by their strong magnetic fields.

\subsection{RM Revolutions fit to intrinsic line maps}
\label{sec:RMR}

\begin{figure}
\includegraphics[trim=2cm 0cm 0.8cm 0.5cm,clip=true,width=\columnwidth]{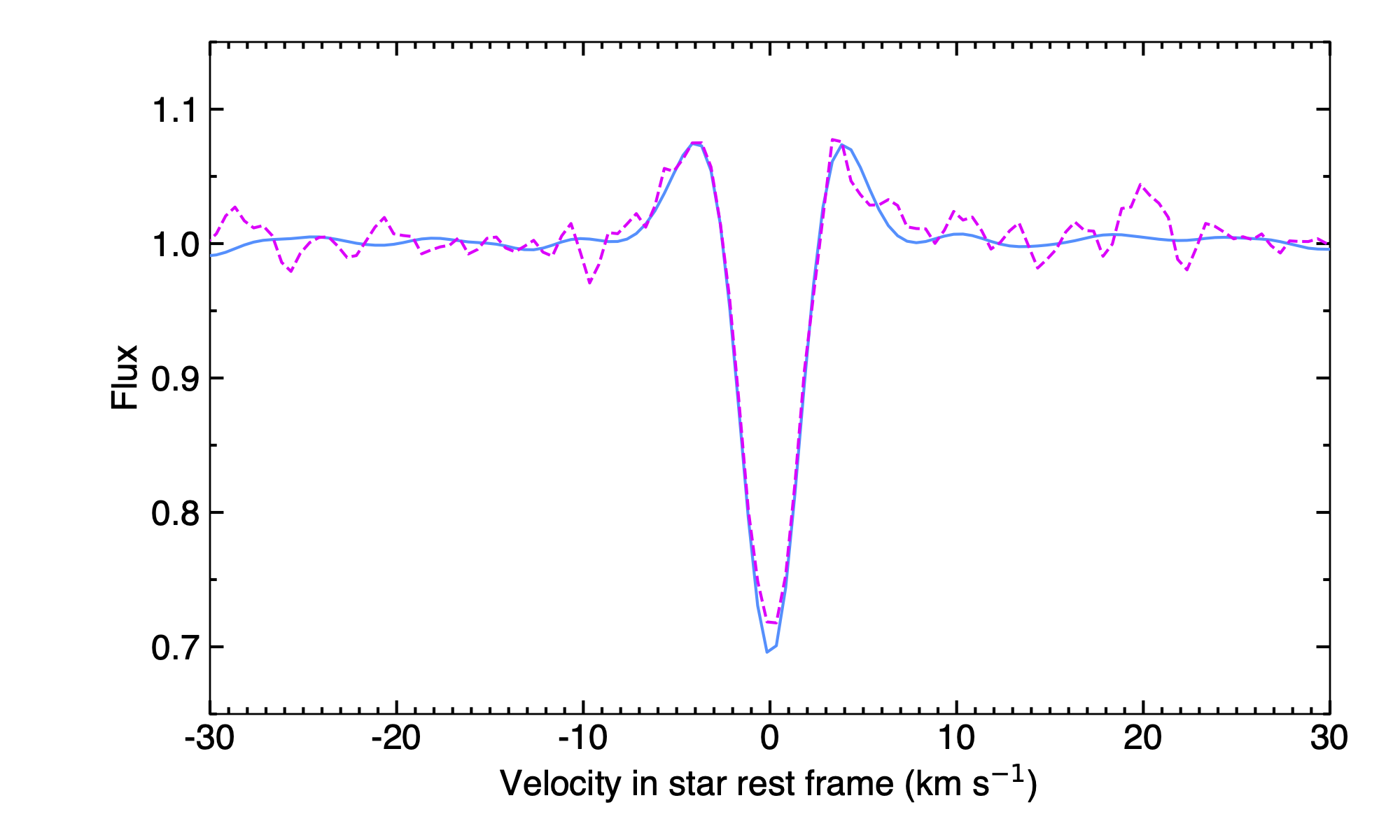}
\centering
\caption[]{Comparison between the master-out CCF$_\mathrm{DI}$ averaged over both ESPRESSO visits (blue), and the average of all CCF$_\mathrm{intr}$ (magenta) after they were shifted to a common rest frame using the best-fit surface RV model for GJ\,436.}
\label{fig:compa_masters}
\end{figure}

The RM Revolutions approach exploits the full information contained in the transit time-series by directly fitting a model of the stellar line to all CCF$_\mathrm{intr}$ simultaneously (details can be found in \citealt{Bourrier2021}). The local stellar line is modeled as a double-Gaussian profile, whose centroids are set by the same surface RV model as in Sect.~\ref{sec:RMR}. Figure~\ref{fig:compa_masters} shows that the master disk-integrated and intrinsic CCFs have the same profile. Comparisons of the Bayesian information criterion (BIC, \citealt{Schwarz1978,Kass1995,Liddle2007}) confirm that it is not justified to model the local stellar line with profiles specific to each visit (BIC = 2276) or common to both visits (BIC = 2259), compared to using the profiles describing the disk-integrated line (BIC = 2239). Properties linking the core and lobe components of the double-Gaussian CCF$_\mathrm{intr}$ model were therefore fixed to those derived for the master CCF$_\mathrm{DI}$ in each visit (Sect.~\ref{sec:DI_fit}). The line shape is therefore controlled by its contrast and FWHM, which are chosen to be constant along the transit chord but specific to each visit (allowing for center-to-limb variations in the local contrast increased the BIC of the fit by 14). The time-series of theoretical stellar lines was convolved with a Gaussian profile of width equivalent to ESPRESSO resolving power, before being fitted to the CCF$_\mathrm{intr}$ in both visits. MCMC jump parameters are the line contrast and FWHM, $\lambda$, and $v$\,sin\,$i_*$. Uniform priors are set on all parameters, and over the same range as for the Reloaded fit for $\lambda$ and $v$\,sin\,$i_*$, between $\mathcal{U}$(0,10)\,km\,s$^{-1}$ for the local FWHM, and between $\mathcal{U}$(0,1) for the local contrast. We ran 60 walkers for 3000 steps, with a burn-in phase of 800 steps. 

\begin{figure}
\includegraphics[trim=0cm 0cm 0cm 0cm,clip=true,width=\columnwidth]{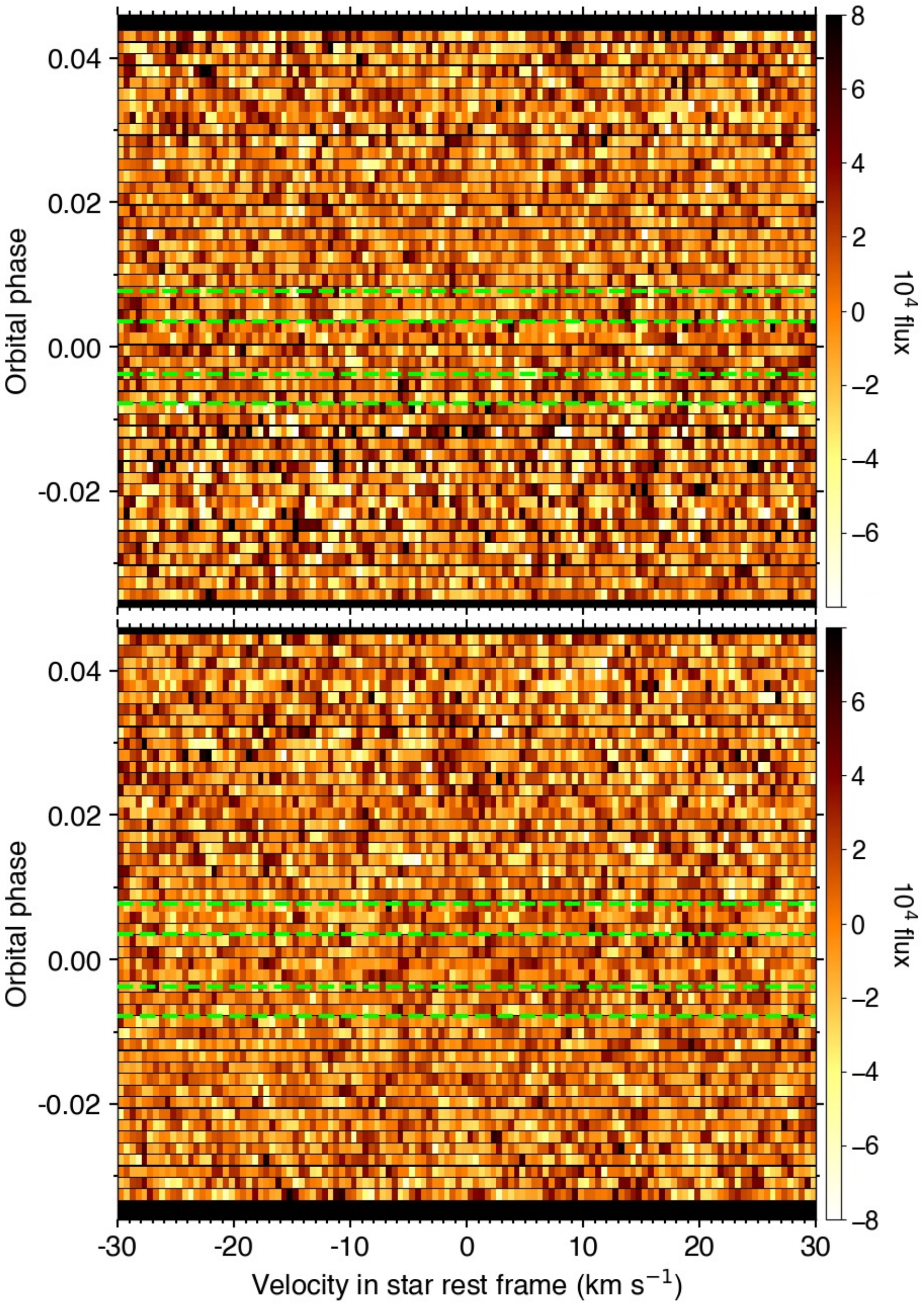}
\centering
\caption[]{Maps of the out-of-transit residuals, and of the in-transit residuals between CCF$_\mathrm{intr}$ and their best-fit RM Revolutions model in ESPRESSO Visit 1 (top panel) and Visit 2 (bottom panel). Transit contacts are shown as green dashed lines. }
\label{fig:Res_map_GJ436b}
\end{figure}

The best-fit line model accurately reproduces the local stellar lines along the transit chord, as can be seen from the residual maps in Fig.~\ref{fig:Res_map_GJ436b}. These maps also highlight the stability of the star over the visits and the overall quality of the data, with no spurious features visible in the out-of-transit residuals. The surface RVs and line properties associated with the best fit are shown in Fig.~\ref{fig:PropLoc_GJ436b}. We derive $v$\,sin\,$i_*$ = 300.5$\stackrel{+45.9}{_{-57.0}}$\,m\,s$^{-1}$ and $\lambda$ = 114.1$\stackrel{+22.8}{_{-17.8}}^{\circ}$. Again these values are consistent within 1$\sigma$ and comparable in terms of precision with those derived from the classical RM and Reloaded analysis (Table~\ref{tab:RM_results}). Varying the mid-transit time and transit depth within their 3$\sigma$ uncertainties and letting the semi-major axis and orbital inclination free to vary with priors set from \citet{Maxted2021} has no impact on the derived values.

\section{Revision of the system architecture.}
\label{sec:glob_fit}

To refine the orbital architecture of the GJ\,436 system  as much as possible we performed a combined RM Revolutions fit to the archival HARPS/HARPS-N and new ESPRESSO data.

\subsection{Reduction of HARPS/HARPS-N data}

One HARPS transit and two HARPS-N transits of GJ\,436b were used by \citet{Bourrier_2018_Nat} to measure its obliquity using the reloaded RM method (we refer to this publication for details). Here we reduced and analyzed these datasets following the same approach as with the ESPRESSO data, albeit with some specificities, which are described below.

Older versions of the HARPS/HARPS-N DRS did not correct for the color effect in M dwarf data, which was therefore corrected for manually by \citet{Bourrier_2018_Nat}. Here we used versions 3.5 and 3.7 of the HARPS and HARPS-N DRS, which apply the color correction 
automatically. The same template as with ESPRESSO was used. All exposures were well corrected for, except for the last one in each HARPS-N night obtained at low S/N because of high airmass and shorter exposure time. These were excluded from our analysis. CCFs were calculated with a step of 0.82\,km\,s$^{-1}$ using the same mask as the ESPRESSO data. 

\begin{figure}
\includegraphics[trim=0cm 0cm 0cm 0cm,clip=true,width=\columnwidth]{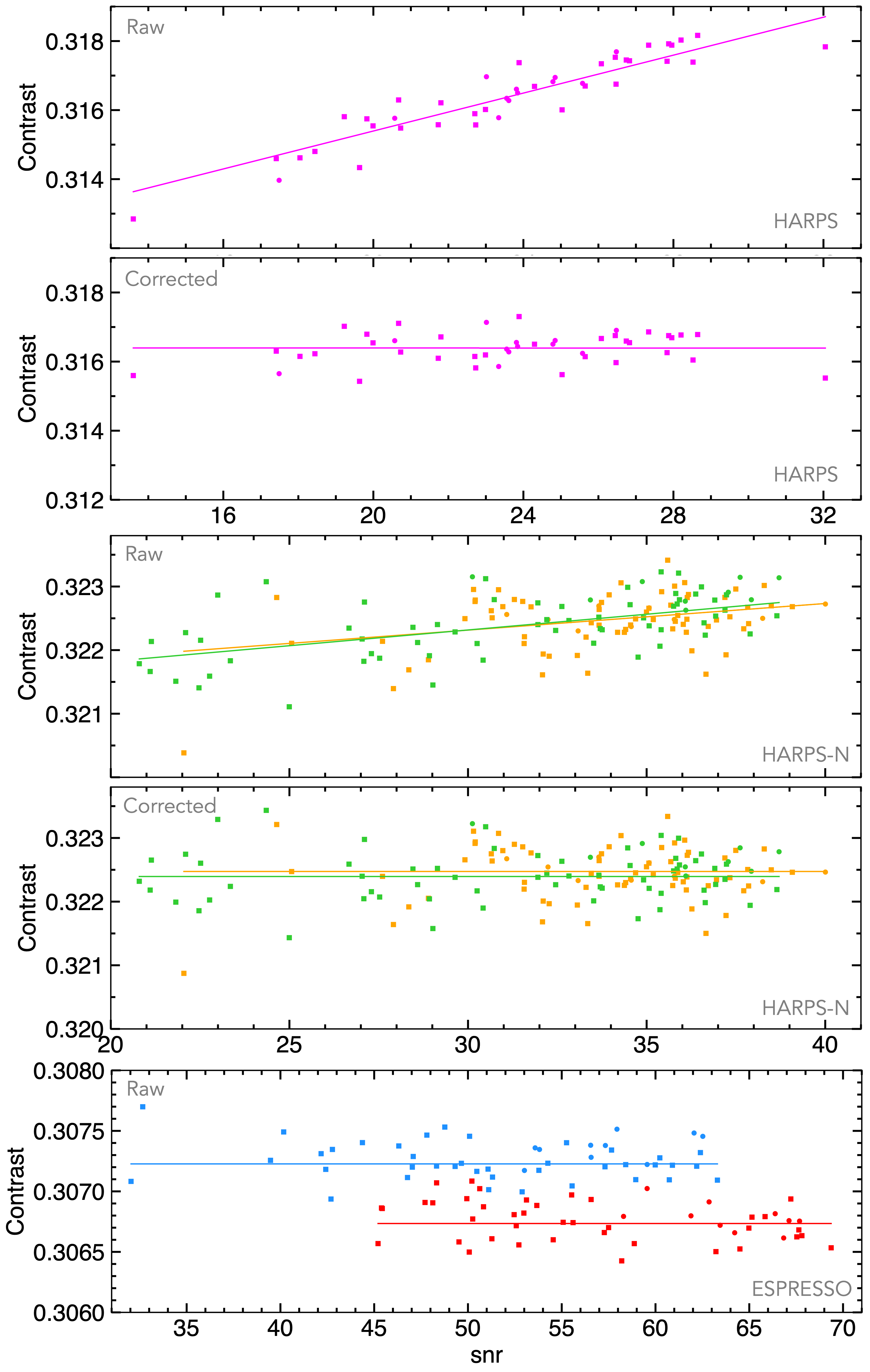}
\centering
\caption[]{Contrast of the HARPS (magenta), HARPS-N (green and orange), and ESPRESSO (blue and red) CCF$_{\rm DI}$. HARPS/HARPS-N data are shown before (top panels) and after (bottom panels) correcting for their correlation with the S/N at 550\,nm. Solid lines show best-fit models to the out-of-transit data (squares; in-transit points are shown as disks). Error bars are not shown for the sake of clarity.  }
\label{fig:C_snr}
\end{figure}

We observed significant correlations between the contrast of the CCF$_{\rm DI}$ and the S/N. These correlations are similar between the HARPS-N visits and particulary strong in the HARPS visit (Fig.~\ref{fig:C_snr}). The origin of these variations is unclear, and we therefore applied an empirical correction obtained by fitting a linear model between the out-of-transit constrast values and the S/N at 550\,nm. This polynomial degree was set by minimizing the BIC of the fit. The correction is applied by temporarily  subtracting CCF$_{\rm DI}$ for their continuum flux (so that their continuum is set to zero), ``stretching'' them using the linear contrast model (so that their contrast is corrected to the expected value), and then adding back the continuum flux. We note that no equivalent correlations are observed in the ESPRESSO data. 

\subsection{Global Revolutions fit}

CCF$_{\rm intr}$ were extracted in the same way as for the ESPRESSO data. GJ\,436 is occulted by the planet in 12 exposures in the HARPS data, and in 9 exposures in the HARPS-N data. Applying the same analysis as in Sect.~\ref{sec:extra} (see Fig.~\ref{fig:PDF_HARPS_indivexp} and \ref{fig:PDF_HARPN_indivexp}), we exclude exposures that display uniform RV PDFs and/or contrast and FWHM PDFs that are consistent with null values, which is an indication that the S/N of their CCF$_{\rm intr}$ is too low to detect the stellar line. This concerns exposures at the limbs of the star: the three first and three last ones in the HARPS transit, and the first and last ones in the HARPS-N transits. 

\begin{figure}
\includegraphics[trim=1.5cm 0.2cm 0.7cm 4.5cm,clip=true,width=\columnwidth]{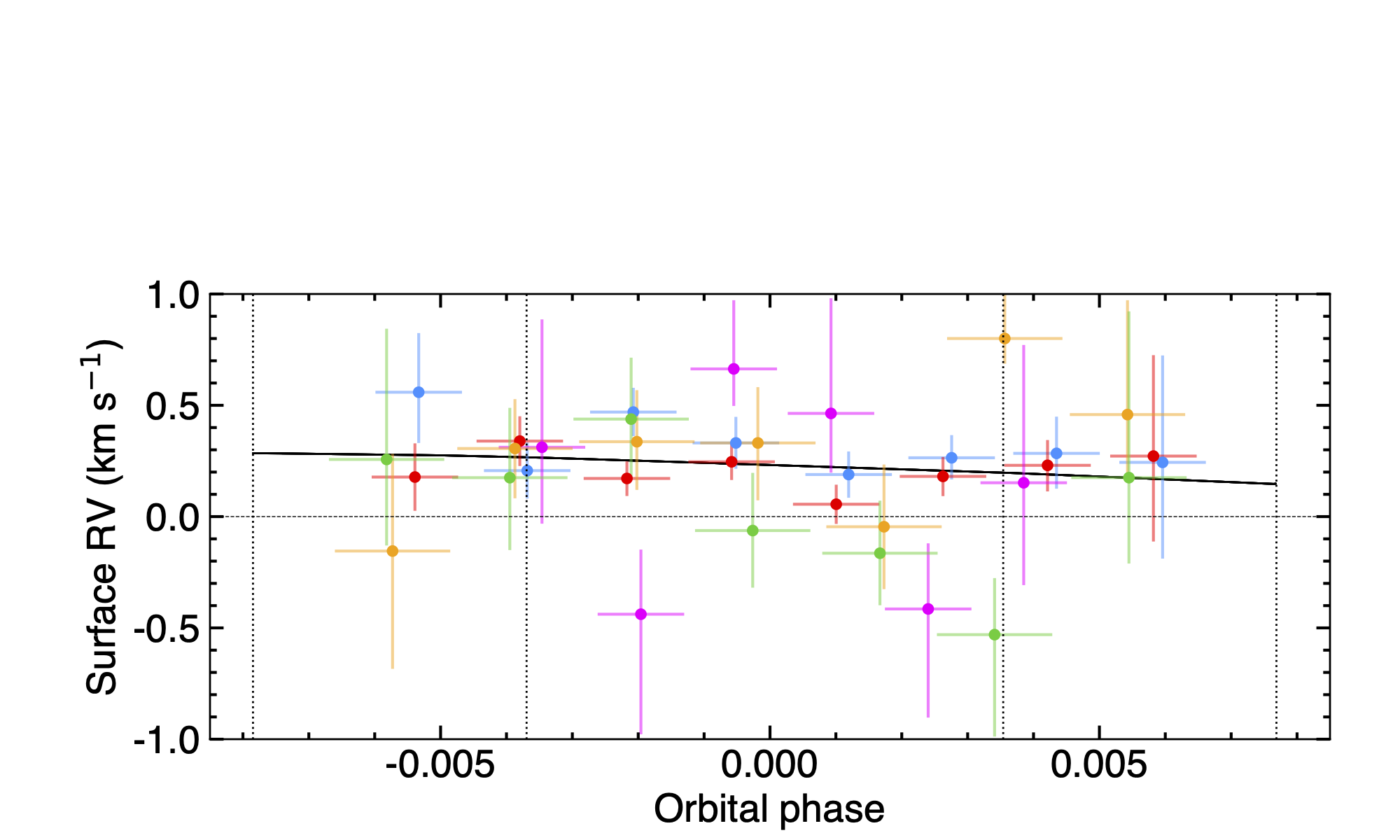}
\centering
\caption[]{Stellar surface RVs of the regions occulted by GJ\,436b in ESPRESSO visits 1 (blue) and 2 (red), HARPS Visit A (magenta), and HARPS-N visits B (orange) and C (green) derived from the fits to individual exposures (same details as Fig.~\ref{fig:PropLoc_GJ436b}). The solid black curve is the best-fit model derived from the joint Revolutions fit to all visits.}
\label{fig:RVsurf_GJ436b_all}
\end{figure}

We used the RM Revolutions technique to fit the five visits together, using the same procedure and settings as in Sect.~\ref{sec:RMR}. The double-Gaussian profile of the local line model is set to the template derived for the disk-integrated line in each visit, with the contrast and FWHM of the line let free to vary in each visit. Theoretical line models are convolved with the instrumental response of each instrument before they are compared with the data. Figure~\ref{fig:RVsurf_GJ436b_all} shows the surface RVs derived from the individual analysis of each exposure (as in Sect.~\ref{sec:extra}), along with the best-fit model derived from the best RM Revolutions fit. We derive $v$\,sin\,$i_*$ = 293.5$\stackrel{+43.7}{_{-52.2}}$\,m\,s$^{-1}$ and $\lambda$ = 113.5$\stackrel{+23.3}{_{-17.3}}^{\circ}$. The correction applied to the HARPS/HARPS-N CCF$_{\rm DI}$ improved the stability of the CCF$_{\rm intr}$ contrast, but has negligible impact on these results. 

We note that combining the HARPS/HARPS-N and ESPRESSO data does not substantially improve the precision on the derived properties (Table~\ref{tab:RM_results}). To understand why, we performed a RM Revolutions fit on the HARPS/HARPS-N data alone, which yielded $v$\,sin\,$i_*$ = 290$\stackrel{+128}{_{-148}}$\,m\,s$^{-1}$ and $\lambda$ = 102$\stackrel{+57}{_{-50}}^{\circ}$. Those results remain consistent within 1$\sigma$ with those obtained by \citet{Bourrier_2018_Nat} but $v$\,sin\,$i_*$ is shifted to lower values while $\lambda$ is shifted to larger values, with both properties having larger uncertainties. First, this explains why the ESPRESSO dataset dominates the global fit of the five visits. Also, these differences with \citet{Bourrier_2018_Nat} could be due to biases in the reloaded RM technique these authors applied to the HARPS/HARPS-N datasets. Indeed, Fig.~\ref{fig:RVsurf_GJ436b_all} shows that the surface RVs have asymmetrical uncertainty ranges resulting from non-Gaussian PDFs from the MCMC fits to the HARPS/HARPS-N CCF$_{\rm intr}$. In such a case, the least-square minimization used to fit CCF$_{\rm intr}$ in the traditional reloaded RM approach yields biased estimates for surface RVs and their uncertainties, which then bias the properties derived by fitting those RV series. This highlights the utility of the RM Revolutions technique in fitting the series of CCF$_{\rm intr}$  directly using a MCMC approach rather than fitting their properties in a two-stage approach as with the reloaded RM technique.

\begin{figure}
\includegraphics[trim=0cm 3cm 0cm 0cm,clip=true,width=\columnwidth]{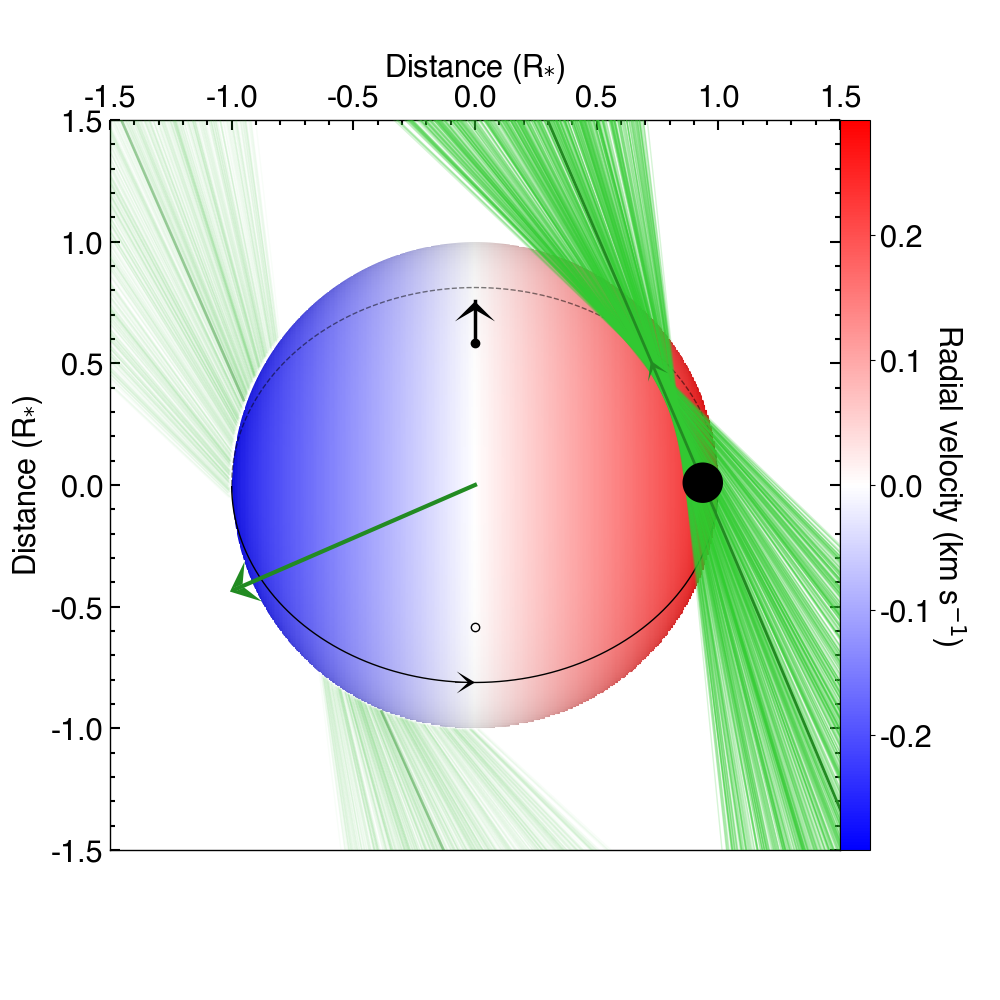}
\centering
\caption[]{Projection of GJ\,436 in the plane of the sky for the best-fit orbital architecture, in the configuration where the stellar spin (shown as a black arrow extending from the north pole) is pointing toward the Earth. The south pole is shown as en empty disk. The stellar equator is represented as a solid (dashed) black line on the visible (hidden) stellar hemisphere. The stellar disk is colored as a function of its surface RV field. The normal to the orbital plane of GJ\,436b is shown as a green arrow. The thick green solid curve represents the best-fit orbital trajectory. The thin lines surrounding it show orbits obtained for orbital inclination, semi-major axis, and sky-projected obliquity values drawn randomly within 1$\sigma$ from their probability distributions. The star, planet (black disk), and orbit are to scale.}
\label{fig:System_view}
\end{figure}

Our measurement of $v$\,sin\,$i_*$ is the most precise to date of all those obtained by analysis of the RM effect, with only five other exoplanet host stars having measurements more precise than 50\,m\,s$^{-1}$ (based on the NASA exoplanet archive). Following the approach of \citet{Bourrier_2018_Nat}, we combined our PDF for $v$\,sin\,$i_*$ with Gaussian PDFs for the stellar radius (0.438$\pm$0.013\,R$_{\Sun}$, \citealt{Maxted2021}) and rotation period (44.09$\pm$0.08\,d, \citealt{Bourrier_2018_Nat}) to build a PDF for the stellar inclination, revising its value to $i_\star$ = 35.7$\stackrel{+5.9}{_{-7.6}}^{\circ}$ or 144.2$\stackrel{+7.6}{_{-5.9}}^{\circ}$. The PDFs for the true 3D obliquity of GJ\,436b associated with these two degenerate configurations have similar widths and medians\footnote{For transiting planets, the 3D obliquity $\Psi$ = $\arccos(\sin(i_\star) \cos(\lambda) \sin(i_p) + \cos(i_\star) \cos(i_p))$ approximates as $\arccos(\sin(i_\star) \cos(\lambda))$} and we averaged them to derive $\Psi$ = 103.2$\stackrel{+12.8}{_{-11.5}}^{\circ}$ (Table~\ref{tab:sys_prop}). The corresponding architecture for the GJ\,436 system is shown in Fig.~\ref{fig:System_view}.

\section{Conclusion}
\label{sec:conclu}

The classical, Reloaded RM, and RM Revolutions analysis of two transits of GJ\,436b observed with ESPRESSO yield similar results for the sky-projected obliquity $\lambda$ and stellar rotational velocity $v$\,sin\,$i_*$, which are also consistent with the values obtained by \citet{Bourrier_2018_Nat} (Table~\ref{tab:RM_results}). This  unambiguously confirms the conclusions  of these latter authors that GJ\,436b is on a polar orbit around its slow-rotating host. 

The three analysis techniques yield unbiased results on $\lambda$ and $v$\,sin\,$i_*$ with comparable precision when the occulted stellar lines are detectable in each exposure, are constant across the transit chord, and solid-body rotation dominates their Doppler shift, as is the case for GJ\,436b. 

The smaller uncertainties provided by the classical RM fit of GJ\,436b likely arise from the underlying assumption of a constant stellar line profile, which removes freedom to the fit compared to the other techniques. However, this assumption is only validated by the independent analysis of the local line profiles in the Reloaded RM and RM Revolutions techniques. In a general case, the classical RM approach should be used with caution because it does not allow the occulted stellar lines to be extracted and studied, making its results prone to bias if these lines vary.

The comparable precision between the Reloaded RM and RM Revolutions fits of GJ\,436b likely arises from the detectability of the occulted stellar lines in nearly all exposures. Indeed, the reloaded RM technique relies on the possibility to derive the RVs of planet-occulted regions by fitting the occulted stellar line in individual exposures. This technique thus loses precision as planets get smaller/host stars fainter and fewer lines can be detected, up to the point where it fails entirely. This was the case with the super-Earth HD\,3167b, whose signal could only be detected through the RM Revolutions technique (\citealt{Bourrier2021}).

While the classical, Reloaded RM, and RM Revolutions techniques yield comparable results in ideal cases, the latter is required to fully exploit a RM transit dataset, and assess and avoid the presence of possible biases due to line shape variations, and we encourage its use to measure obliquities from all types of planetary systems.


\begin{acknowledgements}
We are grateful for a constructive review from the referee, which led us to go deeper into our analyses. The authors acknowledge the ESPRESSO project team for its effort and dedication in building the ESPRESSO instrument. This work has been carried out in the frame of the National Centre for Competence in Research PlanetS supported by the Swiss National Science Foundation (SNSF). The authors acknowledge the financial support of the SNSF. This project has received funding from the European Research Council (ERC) under the European Union's Horizon 2020 research and innovation programme (project {\sc Spice Dune}, grant agreement No 947634; project {\sc Four Aces}; grant agreement No 724427). This work was supported by FCT - Funda\c{c}\~{a}o para a Ci\^encias e a Tecnologia through national funds and by FEDER through COMPETE2020 - Programa Operacional Competitividade e Internacionaliza\c{c}\~{a}o by these grants: UID/FIS/04434/2019; UIDB/04434/2020; UIDP/04434/2020; PTDC/FIS-AST/32113/2017 \& POCI-01-0145-FEDER-032113; PTDC/FIS-AST/28953/2017 \& POCI-01-0145-FEDER-028953; PTDC/FIS-AST/28987/2017 \& POCI-01-0145-FEDER-028987. M.R.Z.O. acknowledges funding from the Spanish Ministery of Science and Innovation through project PID2019-109522GB-C51. V.A. acknowledges the support from FCT through Investigador FCT contracts nr. IF/00650/2015/CP1273/CT0001. NJN acknowledges support form the following projects: UIDB/04434/2020 \& UIDP/04434/2020, CERN/FIS-PAR/0037/2019, PTDC/FIS-OUT/29048/2017, COMPETE2020: POCI-01-0145-FEDER-028987 \& FCT: PTDC/FIS-AST/28987/2017. ASM, JIGH, CAP, and RR acknowledge financial support from the Spanish Ministry of Science and Innovation (MICINN) project AYA2017-86389-P. JIGH also acknowledges financial support from the Spanish MICINN under 2013 Ram\'on y Cajal program RYC-2013-14875. R. A. is a Trottier Postdoctoral Fellow and acknowledges support from the Trottier Family Foundation. This work was supported in part through a grant from FRQNT. M.R.Z.O. acknoledges financial support from project PID2019-109522GB-C51 from the Spanish Ministerio de Ciencia e Innovaci\'on. DE acknowledges financial support from the Swiss National Science Foundation for project 200021\_200726. S.G.S acknowledges the support from FCT through Investigador FCT contract nr. CEECIND/00826/2018 and  POPH/FSE (EC).
\end{acknowledgements}

\bibliographystyle{aa} 
\bibliography{biblio} 

\clearpage

\begin{appendix}

\section{PDFs of the fits to individual exposures}
\label{apn:corr_cont}

\begin{figure*}
\begin{minipage}[h!]{\textwidth}
\includegraphics[trim=0cm 0cm 0cm 0cm,clip=true,width=\columnwidth]{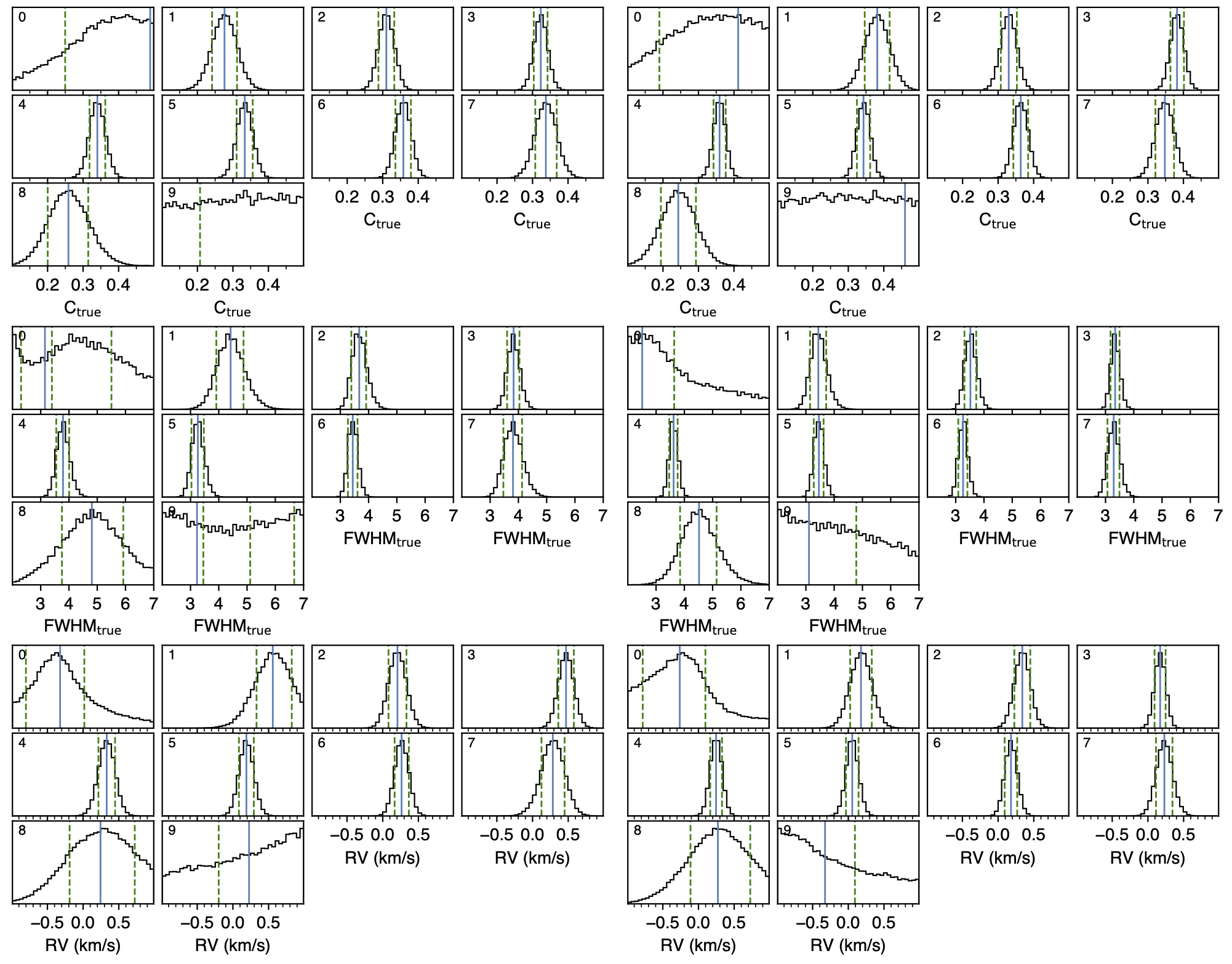}
\centering
\end{minipage}
\caption[]{PDFs of the contrast (upper panels), FWHM (middle panels), and RVs (lower panels) of the Double-Gaussian line model fitted to individual CCF$_\mathrm{intr}$ in ESPRESSO Visits 1 (left) and 2 (right). Blue lines indicate the PDF median values, with dashed green lines showing the 1$\sigma$ highest density intervals (multiple in case of multimodal PDFs). In-transit exposure indexes are shown in each subplot.}
\label{fig:PDF_ESPRESSO_indivexp}
\end{figure*}

\begin{figure*}
\begin{minipage}[h!]{\textwidth}
\includegraphics[trim=0cm 0cm 0cm 0cm,clip=true,width=\columnwidth]{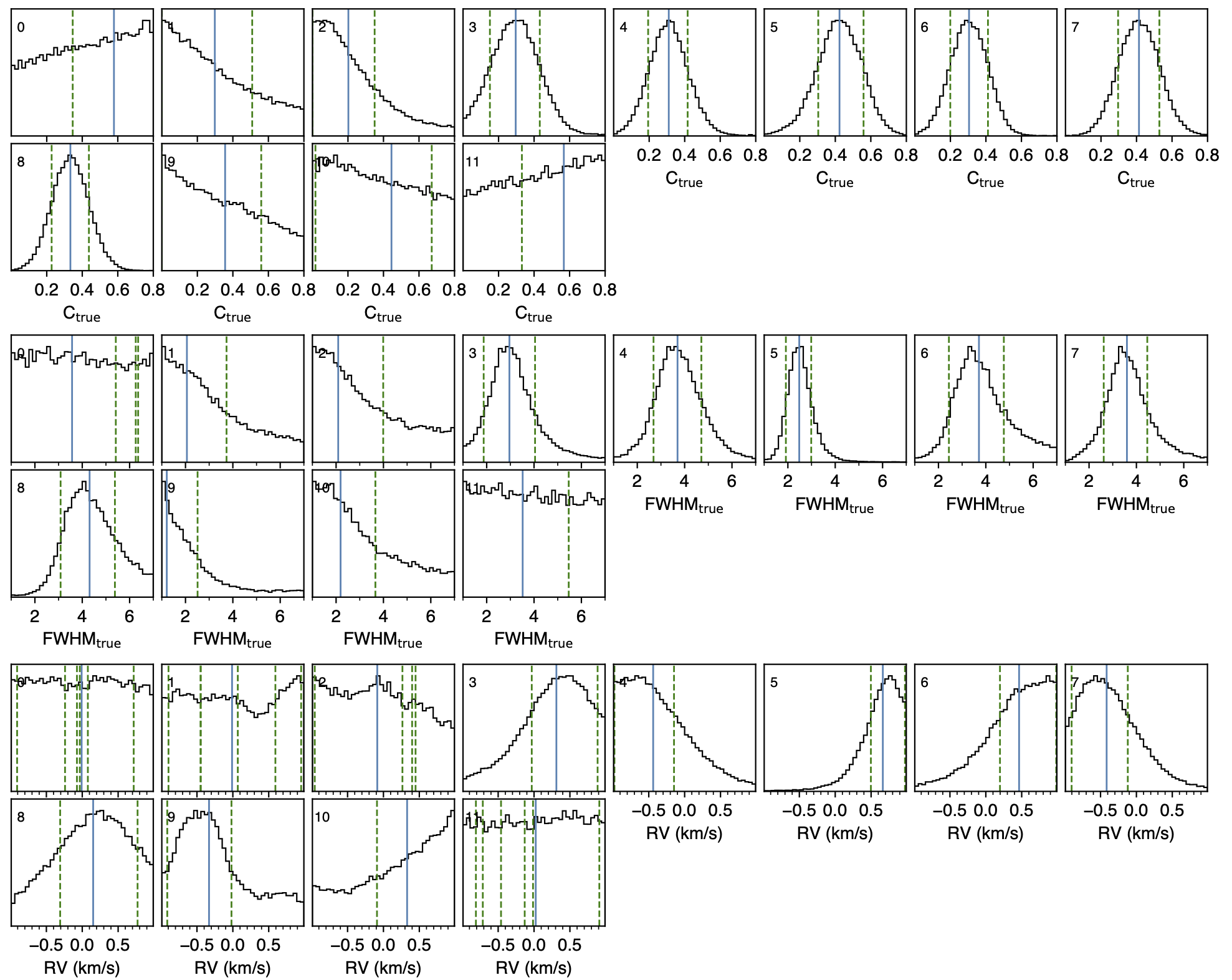}
\centering
\end{minipage}
\caption[]{PDFs of the CCF$_\mathrm{intr}$ properties for the HARPS transit (same details as Fig.~\ref{fig:PDF_ESPRESSO_indivexp}).}
\label{fig:PDF_HARPS_indivexp}
\end{figure*}

\begin{figure*}
\begin{minipage}[h!]{\textwidth}
\includegraphics[trim=0cm 0cm 0cm 0cm,clip=true,width=\columnwidth]{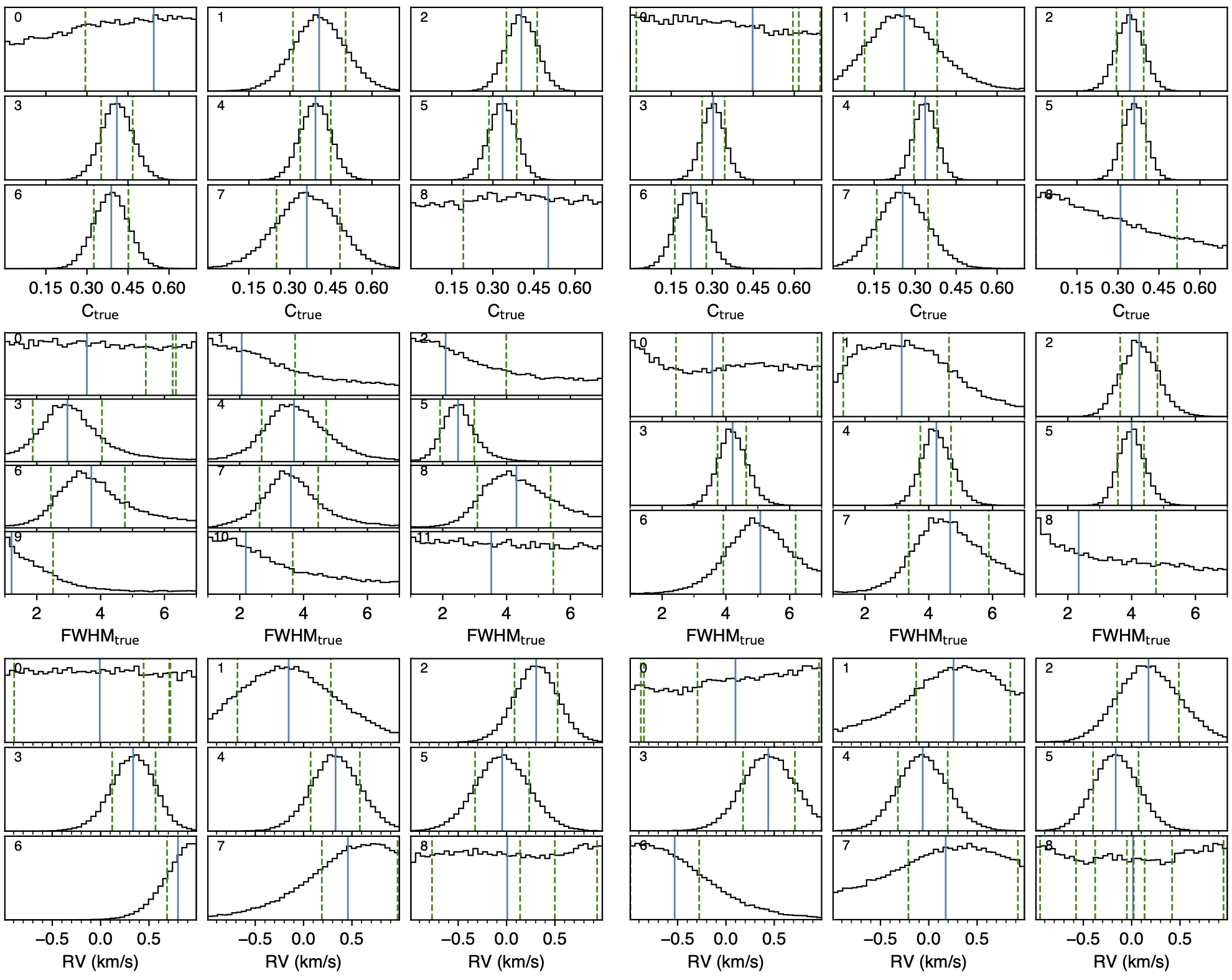}
\centering
\end{minipage}
\caption[]{PDFs of the CCF$_\mathrm{intr}$ properties for the first (left) and second (right) HARPS-N transit (same details as Fig.~\ref{fig:PDF_ESPRESSO_indivexp}).}
\label{fig:PDF_HARPN_indivexp}
\end{figure*}

\end{appendix}

\end{document}